\documentstyle[12pt,epsf]{article} 
\def\pri{^{\, \prime}}

\def\ra{\rightarrow}

\def\prd#1{{\em Phys.~Rev.}~{\bf D#1}\ }
\def\prl#1{{\em Phys.~Rev.~Lett.}~{\bf #1}\ } 
\def\plett#1{{\em Phys.~Lett.}~{\bf #1B}\ } 
\def\np#1{{\em Nucl.~Phys.}~{\bf B#1}\ }
\def\deg{\ifmmode{^{\circ}}\else ${^{\circ}}$\fi} 
\def\ni#1{\noindent$(#1)\quad$}

\def\bi{\begin{itemize}}
\def\ei{\end{itemize}} 
\def\ed{\end{document}}
\def\be{\begin{equation}} 
\def\ee{\end{equation}}
\def\bea{\begin{eqnarray}}
\def\eea{\end{eqnarray}}
\def\labeq#1{\label{eq:#1}} 
\def\vev#1{\left<{#1}\right>} 
\def\lm{\ifmmode{\lambda}\else $\lambda$\fi}
\def\lmp{\ifmmode{\lambda\pri}\else $\lambda\pri$\fi}

\def\mp{\ifmmode{M_P}\else $M_P$\fi}
\def\mgt{\ifmmode{M_{\rm GUT}}\else $M_{\rm GUT}$\fi}
\def\tfrac#1#2{{\textstyle\frac{#1}{#2}}}
\def\half{\frac{1}{2}} 

\def\thalf{\tfrac{1}{2}} 

\def\tthird{\tfrac{1}{3}}

\def\gev{\ \mbox{GeV}}

\def\ms{m_{3/2}}
\def\mo{m_0}
\def\mos{\mo^2}
\def\msq{m_S^2}
\def\su#1{\ifmmode{\mbox{SU}(#1)}\else SU($#1$)\fi}

\def\eb{\end{thebibliography}}

\def\req#1{(\ref{eq:#1})}
\def\eq#1{Eq.~(\ref{eq:#1})}
\def\tr{\ifmmode{\mbox{Tr}}\else Tr\fi} 
\def\bb{\bibitem}
\def\vvs{\ifmmode{\vev{S}}\else $\vev{S}$\fi}
\def\na{{\cal N}_A}
\def\sa{\sum_{a=1}^{\na}}
\def\mg{\ifmmode{\widetilde M}\else $\widetilde M$\fi}
\def\sp{16\pi^2}
\def\ddt#1{\sp\ \frac{d#1}{dt}}
\def\ct{C_2(adj)}
\def\gct{g^2\ct}
\def\gn{g^2N}
\def\nf{N\pri}
\def\ap{A\pri}
\def\aps{A^{\, \prime 2}}
\def\nn{\nonumber}
\def\ma{m_A^2}
\def\map{m_{\ap}^2}
\def\abs#1{\left|#1\right|}
\def\ka{K\"ahler}
\def\js{j^*}
\def\pd#1#2{\frac{\partial #1}{\partial #2}}
\begin{document}
\begin{titlepage} 
\begin{flushright}  {\sl NUB-3154/97-Th}\\  
{\sl Jan 1997}\\  
hep-ph/9701373
\end{flushright} 
\vskip 0.5in 
\begin{center}  
{\Large\bf From Planck to GUT via Dimensional 
Transmutation}\\[.5in]   
{Haim Goldberg}\\[.1in]  
{\sl Department of Physics}\\   
{\sl Northeastern University}\\ 
{\sl Boston, MA 02115}  
\end{center} 
\vskip 0.4in
\begin{abstract}Consider a gauge singlet superfield $S$ 
coupled to a pair
of adjoint fields in a SUSY-GUT. If the tree-level vacuum is flat in $S,$
the vev $\vvs$ which defines the GUT scale will be determined 
via dimensional transmutation at a scale $M$
where the soft-breaking (mass)$^2$ vanishes as a result of running from
$\mp=(8\pi G_N)^{-1/2}.$ Because of the large number of adjoint fields 
$\na$ coupled to $S,$
one finds that $M$ can be  generically close to $\mgt=2\times 10^{16}\gev:$
\[ 
M\simeq \mp\exp[-16\pi^2 \log(3/2)/(\na+4)\lambda^2]\ ,\]
where $\lambda$ is a
Yukawa $\sim 0.7.$  This work examines the symmetries and dynamical
constraints required in a SUSY-GUT in order that the desired
flatness in $S$ is achieved, and that this flatness may survive  in a
supergravity framework. 

\end{abstract} 
\end{titlepage} 
\setcounter{page}{2} 
\section{Introduction}
The realization of a cogent supersymmetric grand unified theory (SUSY-GUT)
has constituted an important goal in particle theory for well over a decade.
Several impediments to achieving such a goal were already evident in the
earliest papers on the subject \cite{DG,Sakai}. In \su{5}, these consist
of the lack of a mechanism within  the theory to $(a)$ lift the degeneracy of
(supersymmetric) \su{5}-, \su{4}$\times$U(1)-, and
\su{3}$\times$\su{2}$\times$U(1)-invariant ground states and $(b)$ implement a
hierarchical splitting  of the massless Higgs doublets from the massive
triplets (D/T splitting) in the ${\bf 5}+\overline{\bf 5}$ representations. The
lifting of the degeneracy is generally ascribed to the perturbation of the
various ground states by the soft-breaking terms \cite{DG,dragon},
while the doublet-triplet splitting can be effected through a judicious 
choice of the Higgs sector \cite{mdm}. 

When the SUSY-GUT is considered in the context of string theory, an additional
problem emerges -- the origin of the GUT scale as distinct from the Planck or
string scale \cite{dienes}.  Generally speaking, the construction of the
superpotential in string theory does not in any obvious way allow for the
introduction of an additional scale $M,$ as is manifest in the \su{5}\
superpotential
\be
W=M\ \tr\ A^2 + \lambda\tr\ A^3\ \ ,
\labeq{dg}
\ee
where $A$ is the \su{5}\ adjoint chiral superfield. It is $M$ that sets the 
scale
$\mgt\simeq 10^{16}-10^{17}\ \gev$ of \su{5}-breaking, and there is no obvious
relation between $\mgt$ and  the Planck scale 
$\mp=(8\pi G_N)^{-1/2}=2.44\times 10^{18}\ \gev$\ \cite{dienes}.

An obvious possibility is to consider, instead of \req{dg}
\be
W\pri = \lm\ S\ \tr\ A^2 + \lmp\ \tr\ A^3 + w(S)\ \ ,
\labeq{wpr}
\ee
where $S$ is an \su{5}\ singlet chiral superfield, and search for a mechanism
which yields $\vvs\ne 0,$ and $M=\lm\vvs\sim\mgt.$ It is clear that, without
additional input, 
\req{wpr} as it stands is not viable, since any $Z_N$ (or U(1))
invariance invoked in order to restrict $w(S)$ will allow $w(S)\sim S^3,$
which, for arbitrary Yukawas, forces $\vvs=0$ at tree level .

In this work, I will examine the possibility of generating $\mgt$ from
\mp\ through radiative corrections in the soft-breaking sector, with a 
resulting
dimensional transmutation \cite{dmtr} at the scale $M\simeq
\mgt.$ This will turn out to be possible, perhaps even inevitable, under
certain simple, well-defined constraints placed on the superpotential. These
constraints serve to  effect the necessary flatness of the effective
potential near the origin of $S,$ so that the minimum is free to wander off to
the point of dimensional transmutation, $\vvs\sim \mgt.$

\section{Toy Model}
To illustrate some of the salient points, I will begin with the case of only
two visible sector superfields, the singlet $S$ and an adjoint $A.$ For the
moment, I will impose a continuous $R$-symmetry where all superfields have
$R$-character $\tthird,$ so that all terms in the superpotential $W$ are
trilinear. In the case of where the GUT is \su{N}\ the most general
superpotential consistent with these requirements is (omitting couplings)
\be
W=S\ \tr\ A^2 + \tr\ A^3 + S^3\ \ .
\labeq{wa}
\ee
In the absence of soft-breaking, the vacuum is given by $\vvs=\vev{A}=0.$ With
soft-breaking, even at a point where the soft-breaking $\msq=0,$  the vevs will
be shifted to
$\vvs\sim\vev{A}\sim
\ms.$ This could be avoided if the $S^3$ term were absent. However, as remarked
in the introduction, any symmetry prohibiting the $S^3$ term will also
forbid the $A^3$ term. Let us accept this for now, so that one can impose a 
$Z_4$
symmetry with charges $q_4(S)=2,\ q_4(A)=1.$ As a consequence of this and the
$R$-symmetry, the superpotential is determined uniquely:
\be
W_0= \lm\ S\ \tr A^2 = \half\lm S \sum_{a=1}^{\na}A_aA_a\ \ ,
\labeq{wb}
\ee
where $\na=N^2-1$ is the dimension of the adjoint of \su{N}. In this toy
model, it is easy also to include SO(10) in the discussion, in which  case
$\na=45.$

At tree level, in the absence of soft-breaking,  the vacuum corresponding to 
\req{wb} is $\vev{A}=0$ with \vvs\ undetermined. I now introduce the
soft-breaking potential
\be
V_{soft}=\msq S^*S +\ma \sa A_a^*A_a+\thalf\lm AS\sa A_aA_a +h.c. +\thalf
\mg\sa \lambda^T_a\lambda_a\ \ ,
\labeq{vsoft}
\ee
where $\lambda_a$ is the adjoint gaugino, and the standard trilinear coupling
parameter $A$ is (hopefully) not to be confused with the adjoint field $A_a.$
This work will focus on the case where the RG evolution of $\msq$ down from
\mp\  leads to its
vanishing  at some scale $Q=M.$ In that case the 1-loop improvement to
the effective potential at scales near $M$ leads to the replacement in
$V_{soft}$ \cite{dmtr}
\be
\msq\ S^*S\longrightarrow   m^{\, \prime 2} S^*S\ln(S^*S/M^2)\ \ ,
\labeq{rep}
\ee
where $m^{\, \prime\ 2}=-\thalf\left.\frac{d\msq}{dt}\right|_{M},\ t
\equiv\ln(\mp/Q).$ The
potential to be minimized is
\be 
V=\abs{\frac{\partial W}{\partial S}}^2 + \abs{\frac{\partial
W}{\partial A_a }}^2 + \thalf m^{\, \prime\ 2} S^*S\ln(S^*S/M^2) + \ma 
A_a^*A_a+\thalf\lm AS\ A_aA_a +h.c.\ \ ,
\labeq{veff}
\ee
where in accordance with $D$-flatness the adjoint field is chosen along one of
the directions of the Cartan subalgebra. There is no sum on $a$ in
\eq{veff}. 

It is a matter of algebra to see that even in the presence of the soft 
breaking,
$V$ is minimized for $\vev{A_a}=0.$ However, $\vvs$ is now determined: one 
obtains
$\vvs=M/\sqrt{\rm e},$ so that although the gauge symmetry remains unbroken, 
the
adjoint field grows a mass $\lm M/\sqrt{\rm e}.$ This is dimensional
transmutation, the breaking of scale invariance due to renormalization effects.
It will now be seen that  for  generic
choices of parameters,  the RG equations will drive $\msq$ 
negative at a scale $M\sim\mgt.$ 

The RG equations for this model are straightforward to obtain:
\bea
\ddt{g}&=&-\left(\sum S_2(R)-3\ct\right)g^3\nn\\
\ddt{\mg}&=&-2\left(\sum S_2(R)-3\ct\right)g^2\ \mg\nn\\
\ddt{\lm}&=&-\thalf\lm\left[(\na+4)\lm^2-8\gct\right]\nn\\
\ddt{A}&=&-\lm\left[(\na+4)\lm^2 A+8\gct\ \mg\right]\nn\\
\ddt{\msq}&=&-\na\lm^2(\msq+2\ma+A^2)\nn\\
\ddt{\ma}&=&-2\lm^2(\msq+2\ma+A^2)+8\gct\ \mg^2\ \ ,
\labeq{rgtoy}
\eea
where $\ct=N$ for \su{N}, 8 for SO(10), and $S_2(R)$ is the Dynkin index of
any gauge-coupled field. As defined previously, $\na$ is the dimension of the
adjoint. Standard initial conditions are imposed on the soft scalar masses:
$\msq(\mp)=m_A^2(\mp)=\mos.$ For simplicity of discussion, I will assume in
all that follows that the quantity $\sum S_2(R)$ is such that the gauge
coupling is essentially constant between \mp\ and the scale $M.$ As a result,
the gaugino mass $\mg$ will also be constant. 

Examination of the evolution equation for $\msq$ in \req{rgtoy} reveals
immediately why dimensional transmutation is likely to occur at  scale $M$ not
far below \mp:
\bi 
\item there is a large factor of $\na$ multiplying the right
hand side: \\[-0.3in]
\item there is no  gaugino contribution serving to retard
the decrease of $\msq$ with momentum scale.\ei
Neither of these
properties characterize the
$m_A^2$ equation, and both are directly tied to  the gauge
singlet nature of $S.$ A simple analytic treatment is heuristic:
take
$A=\mg=0,\
\lm=constant.$ (The latter will be strictly true only at the fixed point.) Then
a simple integration of the last two of Eqs.~\req{rgtoy} gives the solution
\be
\frac{\msq}{\mos}=1-3\left(\frac{\na}{\na+4}\right)\left(1-e^{-\kappa
t}\right)
\ \ ,
\labeq{mssol}
\ee
where $\kappa= \left((\na+4)\lm^2/16\pi^2\right).$ Thus, to a good
approximation,
$\msq=0$ at $t_1=\ln(3/2)/\kappa,$ or
\be
Q_1=M=\mp\ e^{-16\pi^2\log(3/2)/(\na+4)\lm^2}\ \ .
\labeq{q1}
\ee
Because of the large size of $\na+4,$ the evolution to the point of
dimensional transmutation is rapid: from \eq{q1}, $M=\mgt$ in \su{5} for
$\lm\simeq 0.7.$ 

For $A,\ \mg\ne 0,$ some representative numbers can be given. 
With 
$g^2/4\pi=1/24,$ $\ A(\mp)=\mg(\mp)=m_0, $ I find $\msq=0$ at 
$M=\mgt$ for $\lm(\mp)=0.57$ in the case of \su{5}, and
$\lm(\mp)=0.34$ in the case of SO(10). There are sizeable arrays of parameter
space for which $M\simeq \mgt,$ and I will present more detail in the
discussion of a more realistic model. Two points may  be noted before
proceeding:
\bi
\item The GUT scale is triggered by dimensional transmutation in the
soft-breaking sector, but it has no explicit or implicit dependence on $\ms:$
it is essentially given by \eq{q1}.\\[-0.3in]
\item Renormalizability is crucial to the dynamical mechanism proposed here.
Thus it is unclear how to relax the requirement of continuous $R$-invariance so
as to allow higher dimension operators (such as
$(S\tr A^2)^n/\mp^{3(n-1)})$ which respect the $Z_4$ or U(1) symmetry
requirement. Such terms may also vitiate $F$-flatness in $S.$ 
For this paper, I maintain the strict $R$-invariance of the superpotential.\ei

\section{Model with Gauge Symmetry Breaking}

A more realistic model requires some mechanism for the breaking of \su{N}. (The
SO(10) case will receive comment later.) As already noted, simply extending 
the original \su{5} model by letting $M\rightarrow S$ is not possible, since
undesirable $S^3$ terms are then permitted in the superpotential. Instead, it 
is
necessary to introduce a second adjoint $\ap,$ and take as the superpotential
\be
W=2\lm\ S\ \tr\ A\ap + 2\lmp\ \tr\ A^2\ap\ \ ,
\labeq{wreal}
\ee
with the $Z_4$ assignments $(S,A,\ap)=(1,1,2).$ Once more, a continuous
$R$-symmetry is imposed, with $R\mbox{-character}=\tthird$ for all fields,
which will forbid terms such as $M\tr \aps,$ as well as all higher
dimensional operators consistent with the $Z_4$ symmetry. 

At tree level, the vacuum configuration for \su{5} in the direction of the
standard model (the ``24'' direction) is given by 
\be
\vev{A}=(\lm/\lmp)\vvs\ diag(2,2,2,-3,-3),\quad \vev{\ap}=0\ \ ,
\labeq{vacreal}
\ee
with $\vvs$ undetermined.

The soft-breaking
potential is now generalized to  
\bea
V_{soft}&=&\msq S^*S +\sa (m_A^2  A_a^*A_a+m_{\ap}^2 A^{\, \prime *}_a A^{\,
\prime }_a + \lm AS A_a \ap_a + h.c.)\nonumber\\
&& +\thalf\ap\lmp\sum_{a,b,c}^{\na}d^{abc}A_aA_b\ap_c + h.c. +
\thalf\sa\mg\lambda^T_a\lambda_a\ \ ,
\labeq{vsoftreal}
\eea
where the $d^{abc}$ is the  symmetric \su{N} tensor. The RG 
equations for this model are: 
\bea
\ddt{\lm}&=&\thalf\lm\left[(N^2+3)\lm^2+3\nf\lmp^2-8\gn\right]\nn\\
\ddt{\lmp}&=&\thalf\lmp\left[6\lm^2+5\nf\lmp^2-12\gn\right]\nn\\
\ddt{A}&=&2\left[(N^2+1)\lm^2A+\nf\lmp^2\ap+4\gn\ \mg\right]\nn\\
\ddt{\ap}&=&2\left[\lm^2 A +\tfrac{3}{2}\nf\lmp^2\ap
+6\gn\ \mg\right]\nn\\
\ddt{\msq}&=&(N^2-1)\lm^2\left[\msq+2(\ma+\map+\aps)\right]\nn\\
\ddt{\ma}&=&2\left[(\lm^2\msq+(\lm^2+2\nf\lmp^2)\ma+(\lm^2+\nf\lmp^2)\map\right.
\nn\\
&& ~~+\left.\lm^2 A^2+\nf\lmp^2\aps-4\gn\ \mg^2\right]\nn\\
\ddt{\map}&=&2\left[(\lm^2\msq+(\lm^2+\nf\lmp^2)\ma+(\lm^2+\thalf\nf\lmp^2)
\right.
\map\nn\\
&& ~~+\left.\lm^2 A^2+\thalf\nf\lmp^2\aps-4\gn\ \mg^2\right]\ \ ,
\labeq{rgreal}
\eea
where $\nf=(N^2-4)/N.$
Once again, one notes the large $\na=N^2-1$ factor, as well as the absence of
the gaugino mass term on the R.H.S. of the $\msq$ equation in \req{rgreal},
allowing, as in the toy model, a rapid evolution of $\msq$ toward zero. In the
present case, there are  factors of $O(N)$ enhancing the decrease of
$\ma,\ \map$ in descending from \mp. Nevertheless, unless $\mg=0$ and
$\lmp(\mp)\ge 1.5,$ these quantities will not be driven negative in the region
above $10^{16}\ \gev.$

\subsection*{Numerical Study}

In Figure 1, I show some sample ranges of parameters which will give
dimensional transmutation at $M=\mgt$ in \su{5}. The numerical data  are
presented as loci in the
$\lm(\mp)$--$\lmp(\mp)$ space for the four sets of initial conditions
$A(\mp)=\ap(\mp)=(0, m_0),$ and
$\mg(\mp)=(0,
m_0).$ The gauge coupling $g^2/4\pi$ is again fixed at $1/24.$ The required
values of the Yukawa $\lm(\mp)$  are all in the range $0.4-0.6,$ 
showing that the dynamics is effectively controlled by the physics
already present in the toy model of the last section.\clearpage 
\begin{figure}[h]
\begin{center}
\epsfxsize=4.8 in
\epsfysize=4.2 in
\hfil
\epsffile{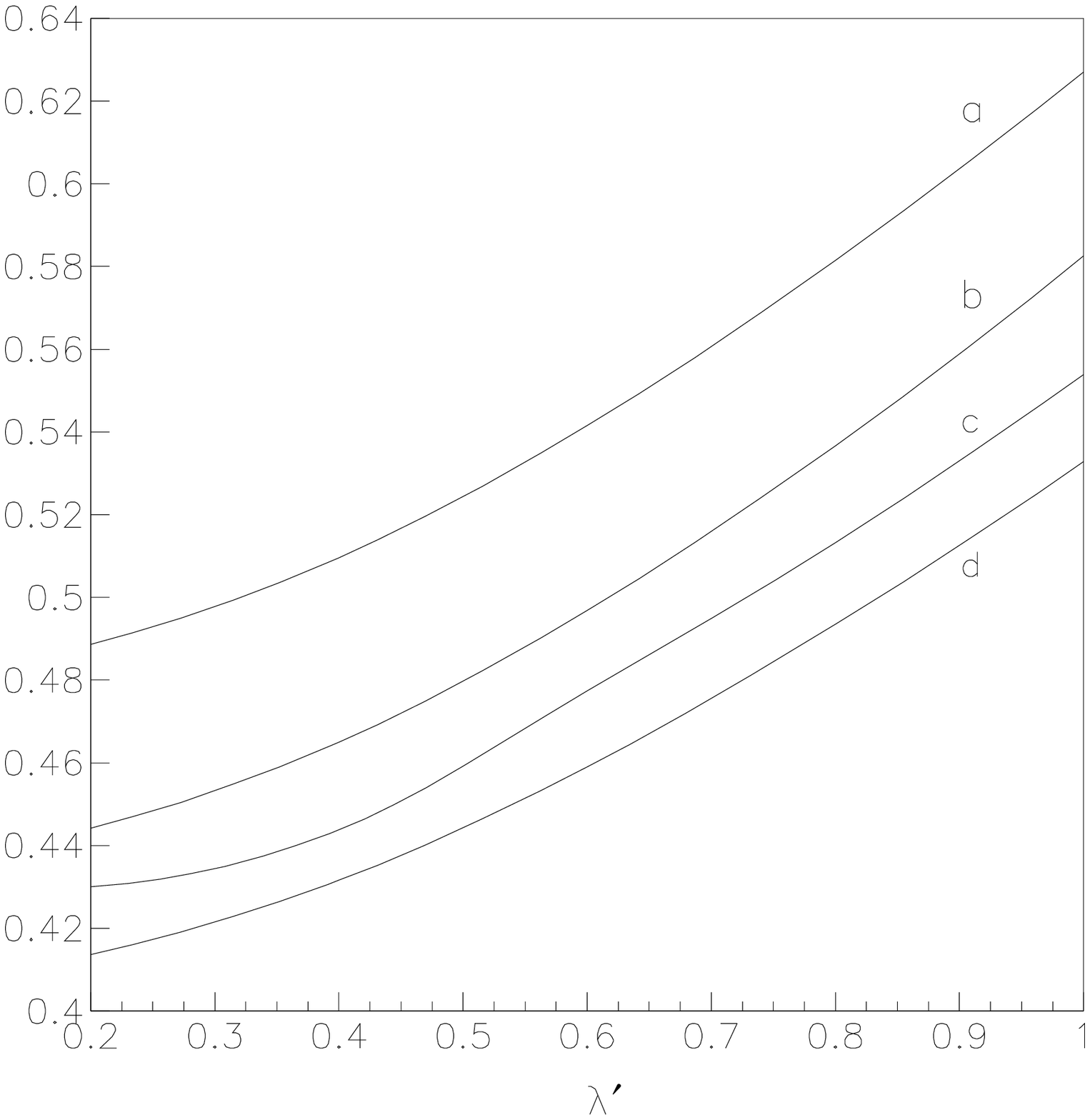}
\hfil
\caption{
Loci in space of Yukawa couplings giving dimensional transmutation at 
$2\times 10^{16}\ \gev,$ for various choices of gaugino mass $\mg(\mp)$ and
trilinear parameters 
$ A(\mp),\ \ap(\mp)$ at the Planck scale.
Curve
$(a): (\mg,A,\ap)=(0.5,\ 0,\ 0)\ m_0;\ 
(b): (\mg,A,\ap)=(0.5,\ 1,\ 1)\ m_0;\  
(c): (\mg,A,\ap)=(1,\ 0,\ 0)\ m_0;\ 
(d): (\mg,A,\ap)=(1,\ 1,\ 1)\ m_0;$
}
\label{dimtj}
\end{center}
\end{figure}

It is straightforward to check the spectrum of this model at scales $Q<M:$
There are 24 Dirac spinors and superpartners with GUT-scale masses, and one
light standard model singlet chiral field. The scalar component of this field
presents a potential Polonyi problem, which will receive some comment in the
concluding section.

\section{Effects of Supergravity}

To what extent are the results presented here stable with respect 
to extension to local
supersymmetry (supergravity)? In a $D$-flat direction, 
the tree-level
potential in local supersymmetry (for the visible sector only) 
corresponding to a superpotential
$W(Z_i)$ is given by \cite{cremmer}
\be
V_{sugra}=e^{K/\mp^2}\left[\left(\pd{W}{Z_i}+\pd{K}{Z_i}\frac{W}{\mp^2}\right)
(K^{-1})_{i\js}\left(\pd{W}{Z_j}+\pd{K}{Z_j}\frac{W}{\mp^2}\right)^*-
3\frac{WW^*}{\mp^2}\right]\ \ ,
\labeq{vsugra}
\ee
where $K(Z_i,Z_i^*)$ is the \ka\ potential and $K^{-1}$ is the inverse of the
matrix $K_{i\js}\equiv
\partial^2K/\partial Z_i\partial Z_j^*.$
The $R$-symmetry restriction to
superpotentials of homogeneous degree 3 implies
\be
\sum_i Z_i\frac{\partial W}{\partial Z_i}=3\ W\ \ .
\labeq{euler}
\ee
For a {\em flat} \ka\ $(K=\sum_iZ_iZ_i^*),$ 
one obtains on inserting \req{euler} into \req{vsugra}
\be
V_{sugra}=\exp(\sum_i \abs{Z_i}^2/\mp^2)\left[\sum_i\abs{\frac{\partial
W}{\partial Z_i}}^2+\left(3+\sum_i
\frac{\abs{Z_i}^2}{\mp^2}\right)\frac{\abs{W}^2}{\mp^2}\right]\ \ .
\labeq{vsugraflat}
\ee

{}From \req{vsugraflat}, we find that $V_{sugra}\ge 0.$ 
Eqs.~\req{vsugraflat} and \req{euler} then ensure that the {\em global
symmetry} condition $\partial W/\partial Z_i=0$ provides a necessary and 
sufficient
condition for the minimum $(V=0)$ to be obtained. From this, it follows 
that if $\vvs$ is not 
determined in the global theory (before  soft-breaking), neither is it
determined in the flat-\ka\ local theory.

What about higher order  terms in the \ka\ potential? For an arbitrary \ka,  
the $R$-symmetry \req{euler} guarantees that
$\partial V_{sugra}/\partial Z_i=V_{sugra}=0$ at the $(S$-flat) field
configuration corresponding to $\partial W/\partial Z_i=0;$  it
does not, of course, guarantee that this field configuration provides a global
minimum for the potential. There is an interesting case where it does: 
consider in the Toy Model of Section 2 a
region of field space where $K=\rho+\thalf a \rho^2/\mp^2,\ \rho=S^*S+A^*A.$ 
This is the U($N$) $(N=2)$ symmetric form suggested by  graviton loop
corrections \cite{lykken}. If $a\ge 0,$ then one can show that  
the minima of the global
and local theories coincide, and $\vvs$ is still undetermined.  
Generally speaking, if $K$ is such as to  destroy $S$-flatness, the vevs will
be moved to $O(\mp),$ the space will become anti-deSitter, and the entire
$R$-symmetry must be dropped in  order to cancel
the resulting $O(\mp^4)$ cosmological constant.\footnote{The hidden sector 
does not, of
course, respect an $R$-symmetry, because of the dual requirements of breaking
SUSY and maintaining a zero cosmological constant.} For
now, I  will just assume that
$K$ behaves in a manner such as to preserve the  vacuum in the $S$-flat
direction, and delay consideration of this point to future study. It must be
noted, however, that  even if $K$  behaves appropriately, the 
local theory  is still 
not renormalizable, so that the dimensional transmutation
requires ignoring the gravitational strength interactions in
obtaining the running of the soft parameters.
 
\section{Summary and Remarks}

\ni{1}In this work, I have demonstrated how the GUT scale \mgt\
could arise through dimensional transmutation at a scale $M$ where the
soft-breaking $\msq$ of a gauge-singlet field $S$ becomes negative and a vev
$\vvs=M/\sqrt{\rm e}$ develops. The scale
$M$ does not depend numerically on the SUSY-breaking scale $\ms,$ and is of
$O(\mgt)$ because the rate of decrease of $\msq$ on descending from \mp\ is
proportional to a large number, the dimension of the adjoint
Higgs representation $A$. At the scale $M$, the adjoint develops a mass $\sim
M,$ and if there is self-coupling, a non-zero vev.

\ni{2}This scenario requires $S$-flatness of the effective potential before
radiative corrections. In this work, this  has been implemented  by  two
symmetries: a continuous
$R$-symmetry which enforces all terms in the superpotential to be trilinear, 
and
a discrete or continuous symmetry which forbids more than a linear dependence 
on
$S$ for any term in the superpotential. Except for possible gravitational
effects
discussed above, the
$R$-symmetry allows the theory to be renormalizable between \mp\ and $M,$
while the additional symmetry keeps $\vvs$ indeterminate at tree
level, allowing dimensional transmutation to take place at the high
scale $\sim \mgt.$

\ni{3}An extension to SO(10) of the second model discussed in this paper
would require a third adjoint (or a symmetric {\bf 54}) in order to create a
trilinear term besides $SA\ap.$ Such an extension, and other non-trivial
modifications (such as those required to accommodate the doublet
triplet splitting) are the subjects of future study. It should be noted that  
every field coupled to the singlet $S$ will tend to
drive the transmutation scale $M$ closer to \mp. This will limit the number
and dimension of such fields.

\ni{4}Many SO(10) models require a set of heavy ${\bf 16} + \overline{\bf
16}$ pairs of superfields to effect the SO(10)$\ra$\su{5} breaking, and 
in order to obtain realistic
low energy Yukawa matrices \cite{ljhall}. By allowing the singlet $S$ to
couple to such pairs, the dimensional transmutation will automatically force
them to grow a mass $M.$ 

\ni{5}The development of a vev for $S$ will break the $Z_4$ (or
U(1)) symmetry used in order to forbid the $S^3$ term. In the $Z_4$ case,
the resulting domain walls can be rendered harmless by a period of post-GUT
inflation. In the U(1) case, the undesirable GUT-scale axion
\cite{sikivie}  is not present if the U(1) is
gauged.  The final cosmological problem is
presented by the scalar component of the field $S\pri=S-\vvs,$ which has a 
mass $\sim
\abs{d\msq/dt}^{1/2}\sim\ms.\;$\footnote{In the two-adjoint model of Section 3, 
the light scalar is a linear combination of $S$ and $\ap_{24}.$} If these 
particles survive to the post-inflation 
era as a long-wavelength
classical field  with amplitude of $S\pri\sim O(\mgt),$ then the 
familiar Polonyi problem
results \cite{coughlan}. After the onset of inflation,
$S\pri$ has a mass $\sim H$ \cite{dfn}, and is localized at $S\pri=0$
\cite{dfn,dvali}. Whether or not it remains localized depends on
its \ka\ couplings to the inflaton and to the fields of the hidden sector
\cite{dinrantho}. Discussion of this awaits a fuller understanding of Planck
scale physics.

\subsection*{Acknowledgement}I would like to thank Stuart Raby for  helpful
comments. This research was supported in part by Grant No
PHY9411546 from the National Science Foundation.
%
%

\begin{thebibliography}{99}
\bb{DG}S.~Dimopoulos and H.~Georgi, \np{193}(1981) 150.  
\bb{Sakai}N.~Sakai,\ {\em Zeit.~Phys.}~{\bf C11}\ (1981) 153.
\bb{dragon}N.~Dragon,\ {\em Zeit.~Phys.}~{\bf C15}\ (1982) 169.
\bb{mdm}S.~Dimopoulos and F.~Wilczek, {\em Proceedings of the Erice Summer
School}, ed. A~Zichichi (1981); 
H.~Georgi, \plett{108}(1982) 283; 
A.~Masiero, D.~V.~Nanopoulos, K.~Tamvakis, and T.~Yanagida, \plett{115}(1982)
380; 
B.~Grinstein, \np{206}(1982) 387.
\bb{dienes}For a comprehensive recent survey of the problem, see Keith~R.~
Dienes, IASSNS-HEP-95-97, hep-th/9602045.
\bb{dmtr}S.~Coleman and E.~Weinberg, \prd{7}(1973) 1883. 
This mechanism was  discussed in the context of radiative
electroweak breaking in the MSSM by J.~Ellis, J.~Hagelin, D.~V.~Nanopoulos,
and K.~Tamvakis, \plett{125}(1983) 275. See also J.~Ellis, A.~B.~Lahanas, 
D.~V.~Nanopoulos, K.~Tamvakis, \plett{134}(1983) 429; C.~Kounnas,
A.~B.~Lahanas, and D.~V.~Nanopoulos, \np{236}(1984) 438. 
An attempt to generate \mgt\ from \mp\ (B.~Gato, J.~L\'eon, and M.~Quir\'os,
\plett{136}(1984) 361) made use of a superpotential
$W\sim \eta \tr A^2, \eta\sim O(\ms),$ which introduces a small mass parameter
in the superpotential. More in the spirit of the present work is a paper by
P.~Moxhay and K.~Yamamoto, \plett{151}(1985) 363, which proposes obtaining the
Peccei-Quinn scale through dimensional transmutation. There have also been
previous efforts to generate $\mgt$ from the weak scale (E.~Witten,
\plett{105}(1981) 267) and from an intermediate SUSY-breaking scale
(S.~Dimopoulos and S.~Raby, \np{219}(1983) 479) via radiative corrections.
\bb{cremmer}E.~Cremmer, S.~Ferrara, L.~Girardello, and A.~van~Proeyen,
\plett{116}(1982) 231; \np{212}(1983) 413; A.~H.~Chamseddine, R.~Arnowitt, and
P.~Nath, \prl{49}(1982) 970; 
E.~Witten and J.~Bagger, \plett{115}(1982) 202; J.~Bagger, \np{211}(1983) 302.
\bb{lykken}L.~J.~Hall, J.~Lykken, and S.~Weinberg, \prd{27}(1983) 2359.
\bb{ljhall}G.~Anderson, S.~Dimopoulos, L.~J.~Hall, S.~Raby, and G.~Starkman,
\prd{49}(1994) 3660; 
K.~S.~Babu and S.~M.~Barr, \prl{75}(1995) 2088.
\bb{sikivie}J.~Preskill, M.~B.~Wise, and F.~Wilczek, \plett{120}(1983) 127; 
L.~F.~Abbott and P.~Sikivie, \plett{120}(1983) 133; 
M.~Dine and W.~Fischler, \plett{120}(1983) 137.
\bb{coughlan}G.~Coughlan, W.~Fischler, E.~Kolb, S.~Raby, and G.~Ross,
\plett{131}(1983) 59; 
J.~Ellis, D.~V.~Nanopoulos, and M.~Quir\'os, \plett{174}(1986); 
 B.~de~Carlos, J.~A.~Casas, F.~Quevedo, and E.~Roulet, \plett{318}(1993) 447.
\bb{dfn}M.~Dine, W.~Fischler, and D.~Nemeschansky, \plett{136}(1984)169.
\bb{dvali}G.~Dvali, hep-ph/9503259, preprint IFUP-TH 09-95, 
\bb{dinrantho}M.~Dine, L.~Randall, and S.~Thomas, \prl{75}(1995) 398
(hep-ph/9503303).
\eb\ed